\documentclass{article}

\usepackage{arxiv}

\usepackage[utf8]{inputenc} % allow utf-8 input
\usepackage[T1]{fontenc}    % use 8-bit T1 fonts
\usepackage{hyperref}       % hyperlinks
\usepackage{url}            % simple URL typesetting
\usepackage{booktabs}       % professional-quality tables
\usepackage{amsfonts}       % blackboard math symbols
\usepackage{nicefrac}       % compact symbols for 1/2, etc.
\usepackage{microtype}      % microtypography
\usepackage{lipsum}
\usepackage{physics}
\usepackage{siunitx}
\usepackage{amsmath}
\usepackage{amsfonts}
\usepackage{amssymb}
\usepackage{gensymb}
\usepackage{tensor}

\usepackage{optidef}
\usepackage{algorithmicx}
\usepackage{algorithm,algpseudocode}

\usepackage{xcolor}
\usepackage{bm}  % for bold symbols 
\usepackage{booktabs}  % better tables
\usepackage{pifont}  % for x mark
\usepackage{graphicx}
\usepackage{hyperref}

\usepackage{pgfplots}
\pgfplotsset{compat=1.15,
	legend style={font=\footnotesize},
}
\usepackage{tikzscale}
\title{A Square-Root Kalman Filter Using Only QR Decompositions}

\author{
  Kevin S. Tracy\thanks{A member of the Robotic Exploration Lab \url{http://roboticexplorationlab.org/}} \\
  The Robotics Institue\\
  Carnegie Mellon University\\
  Pittsburgh, PA 15213 \\
  \texttt{ktracy@cmu.edu} \\
}

\begin{document}
\maketitle

\begin{abstract}
    The Kalman filter operates by storing a Gaussian description of the state estimate in the form of a mean and covariance. Instead of storing and manipulating the covariance matrix directly, a square-root Kalman filter only forms and updates a triangular matrix square root of the covariance matrix. The resulting algorithm is more numerically stable than a traditional Kalman filter, benefiting from double the working precision.  This paper presents a formulation of the square root Kalman filter that leverages the QR decomposition to dramatically simplify the resulting algorithm.
\end{abstract}

% \begin{abstract}
% The Kalman Filter is ubiquitous in state estimation, but can struggle for systems with poor numerical properties. While Kalman Filters operate on the potentially ill-conditioned covariance matrices, square root variants of the Kalman Filter store and manipulate the matrix square root of these matrices. This allows square root Kalman Filters to benefit from double the working precision over the traditional Kalman filter, and significantly less susceptibility to round-off error and ill-conditioning. This paper presents a form of the square root Kalman Filter that leverages QR decompositions to dramatically simplify the resulting algorithm. 
% \end{abstract}

% keywords can be removed
% \keywords{Kalman Filter \and State Estimation \and Stochastic Systems}

\section{Motivation}

Shortly after the introduction of the Kalman Filter \cite{Kalman1960}, work began on implementation for embedded systems that were constrained in both computation and memory \cite{Kaminski1971,Bierman1973}. At the time, Apollo trajectory simulations were being performed at the NASA Ames Research Center (ARC) and MIT on an IBM 704 with 36-bit floating point arithmetic. These computers were adequate for trajectory simulation but began to struggle numerically with computing the covariance update in the Kalman Filter. At that time, the Kalman Filter was unable to be run on the Apollo guidance computer that was only capable of 15-bit fixed-point arithmetic \cite{Grewal2010}.  

James E Potter, a graduate student at MIT, was able to dramatically improve the numerical accuracy of the Apollo guidance computer by storing and manipulating all covariance matrices in the form of their Cholesky factorizations. This meant that only a factored "square root" version of the covariance had to be dealt with, and the full covariance matrix need not be formed. The resulting algorithm was described as \textit{square-root filtering}, and it enabled the implementation of numerically robust filtering on the Apollo guidance computer. This new square-root Kalman filter was able to achieve the same accuracy as the traditional Kalman filter with half as many bits of precision \cite{Grewal2010}.

Many variants of the square-root Kalman filter, and closely related Information filter, have been demonstrated \cite{Kaminski1971}.  This paper presents a significantly simplified square-root Kalman filter by leveraging the QR decomposition to handle the numerically sensitive covariance updates. Because these computations are done by a mature linear algebra library, this filter has even better numerical properties than alternative square-root formulations.  The resulting algorithm is simple and easy to implement, with an example implementation provided at \url{https://github.com/kevin-tracy/QRKalmanFilter}.

% \section{Current State of Square-Root Filtering}

% Since the advances of Potter and the Apollo project, many variations of the square root filter have been proposed. The advancements to square-root Kalman filtering can be put into the following three categories: 
% \begin{itemize}
%     \item Kalman filter vs. Information filter vs. Unscented filter
%     \item Types of matrix factorization used for the covariance square root
%     \item Covariance update given measurements 
% \end{itemize}

\section{Matrix Square Roots}
Of the available interpretations of a matrix square root, this paper will look at those with the following relationship:
\begin{align}
    M &= U^T U ,
\end{align}
where we $U$ is said to be a "square root" of $M$. For positive definite matrices, this square root can be found using a Cholesky factorization, with the resulting square root being upper triangular. The ability to express the square root of a sum of matrices as a function of the individual square roots is required for square-root filtering. This desired operation is expressed using some function $f$ as follows:
\begin{align}
    \sqrt{A + B} = f(\sqrt{A},\sqrt{B}).
\end{align}
In order to achieve this, the sum of matrices can be equivalently written as a product of the following two matrices:
\begin{align}
    A + B = \begin{bmatrix} \sqrt{A}^T & \sqrt{B}^T\end{bmatrix} \begin{bmatrix} \sqrt{A} \\ \sqrt{B}\end{bmatrix}. \label{eq:sqrtm}
\end{align}
With this, an upper triangular square root can be computed with a QR factorization of the matrix on the right:
\begin{align}
    Q,\,R = \operatorname{qr}\bigg( \begin{bmatrix} \sqrt{A} \\ \sqrt{B}\end{bmatrix} \bigg),
\end{align}
where $Q$ is an orthogonal matrix, and $R$ is upper triangular. Substituting the QR decomposition into \eqref{eq:sqrtm}, and using the fact that the orthogonality of Q results in $Q^T = Q^{-1}$, we get the following:
\begin{align}
    A + B &= R^T Q^T Q R, \\
    A + B &= R^T R,
\end{align}
giving us a square root in $R = \sqrt{A + B }$. For ease of notation, this operation of taking the QR decomposition of two vertically concatenated matrices and returning the upper triangular $R$ will be expressed as:

\begin{align}
    \sqrt{A + B} = \operatorname{qr_r}(\sqrt{A},\sqrt{B}).
\end{align}

\section{Square Root Kalman Filter}

A discrete time linear dynamic system with state variable $x\in \mathbb{R}^n$ will be described in state space form:
\begin{align}
    x_{t+1} &= A x_t + B u_t + w_t, \\ w_t &\sim \mathcal{N}(0_n,W),
\end{align}
where $w_t$ is the Gaussian process noise with zero mean and covariance $W\in \mathbb{S}_+^n$. The measurement $y \in \mathbb{R}^m$ will be expressed as the following:
\begin{align}
    y_{t} &= C x_t + v_t, \\ v_t &\sim \mathcal{N}(0_n,V),
\end{align}
where $v_t$ is the Gaussian process noise with zero mean and covariance $V\in \mathbb{S}_+^m$.  Cholesky decompositions of the two noise covariances $W$ and $V$ result in their matrix square roots:
\begin{align}
    \Gamma_W &= \sqrt{W} = \text{cholesky}(W), & 
    \Gamma_V &= \sqrt{V} \,= \text{cholesky}(V).
\end{align}
The Kalman Filter can be divided into three sections: a prediction step where the noise-free dynamics are propagated one time step, an innovation step where the measurement is processed, and an update step where the final mean and covariance is updated. These steps will be written out in their original form in the Kalman Filter, and a formulation for the SQRKF will be derived:
\subsection{Predict Step}
The Kalman Filter maintains an estimate of the state $\mu \in \mathbb{R}^n$, and the covariance $\Sigma \in \mathbb{S}_+^n$. This estimate of the state at time $t$ given the measurement and control histories up to that point is denoted as the following 
\begin{align}
    \mu_{t|t} = \mathbb{E}(x_t | y_{1:t},u_{1:t}),
\end{align}
with covariance
\begin{align}
    \Sigma_{t|t} = \text{cov}(x_t - \mu_{t|t}).
\end{align}
The predict step in the Kalman filter is for estimating the mean and covariance of the state at time $t+1$ given only the measurements and control histories up to time $t$. This is done by propagating the noise-free dynamics forward one timestep:
\begin{align}
    \mu_{t+1|t} &= A \mu_{t|t} + B u_t, \\
    \Sigma_{t+1|t} &= A \Sigma_{t|t} A^T + W.
\end{align}
The SQKF uses the same prediction for the mean, but must adapt the prediction for the covariance to use only the matrix square roots. The square root of the estimate covariance will be written as $F \in \mathbb{R}^{n \times n}$, where $\Sigma_{t|t} = F_{t|t}^TF_{t|t}$.  With this, the square root covariance update is derived:
\begin{align}
\Sigma_{t+1|t} &= A \Sigma_{t|t} A^T + W, \\
    F_{t+1|t}^TF_{t+1|t} &= AF_{t|t}^TF_{t|t} A^T + \Gamma_W^T\Gamma_W, \\ 
     F_{t+1|t}^TF_{t+1|t}  &= \begin{bmatrix} A F_{t|t}^T & \Gamma_W^T\end{bmatrix} \begin{bmatrix}F_{t|t} A^T \\ \Gamma_W\end{bmatrix},\\
    F_{t+1|t} &= qr_r ( F_{t|t} A^T , \Gamma_W).
\end{align}
\subsection{Innovation}
The innovation step in the Kalman filter is the part where the expected measurement is compared to the actual measurement as the innovation $z \in \mathbb{R}^m$, and the covariance of this innovation $S \in \mathbb{S}^m$ is calculated.
\begin{align}
    z &= y_{t+1} - C\mu_{t+1|t}, \\ 
    S &= C \Sigma_{t+1|t} C^T + V.
\end{align}
Similar to the predict step, the vector variable $z$ is left alone, and the innovation covariance expression is factored into square root $G \in \mathbb{R}^{m \times m}$, where $S_t = G_t^T G_t$.
\begin{align}
    S &= C \Sigma_{t+1|t} C^T + V, \\ 
    G^TG &= CF_{t+1|t}^T F_{t+1|t} C^T+ \Gamma_V^T\Gamma_V, \\ 
    G^TG &= \begin{bmatrix} C F_{t+1|t}^T & \Gamma_V^T\end{bmatrix} \begin{bmatrix}F_{t+1|t} C^T \\ \Gamma_V\end{bmatrix}, \\ 
    G &= \operatorname{qr_r}(F_{t+1|t} C^T, \Gamma_V).
\end{align}
The Kalman gain is the linear feedback matrix that maps the innovation to an update in the estimated state. In the normal Kalman Filter, this feedback matrix is calculated as the following:
\begin{align}
    L&= \Sigma_{t+1|t}C^TS^{-1}.
\end{align}
To utilize the standard inputs for a linear solver, the Kalman gain can be transposed in the familiar form:
\begin{align}
    L &= (S^{-1} C \Sigma_{t+1|t})^T.
\end{align}
Since the innovation covariance $S_t$ has already been factored, this can be efficiently solved using forward and backward substitution:
\begin{align}
    L &= [G^{-1}(G^{-T} C) F_{t+1|t}^TF_{t+1|t}]^T.
\end{align}
For languages with a backslash linear solve function that will recognize the triangular structure of the matrix square roots, the Kalman gain can be accomplished as follows:
\begin{align}
    L &= [G \backslash  ((G^T \backslash  C) F_{t+1|t}^TF_{t+1|t}  )]^T.
\end{align}

\subsection{Update Step}
Once the Kalman Gain has been calculated, the mean and covariance can be updated accordingly. 
\begin{align}
    \mu_{t+1|t+1} &= \mu_{t+1|t} + Lz, \\
    \Sigma_{t+1|t+1} &= (I - LC){\Sigma_{t+1|t}}(I-LC)^T + LRL^T. \label{eq:joseph}
\end{align}
The covariance update performed in \eqref{eq:joseph} is the Joseph form update, importantly chosen to maintain positive definiteness of the covariance. This also allows for a factorization of the covariance update:
\begin{align}
    \Sigma_{t+1|t+1} &= (I - LC){\Sigma_{t+1|t}}(I-LC)^T + LRL^T, \\
    F_{t+1|t+1} ^TF_{t+1|t+1}  &= (I - LC){F_{t+1|t}^T F_{t+1|t}}(I-LC)^T + L\Gamma_V^T\Gamma_VL^T,\\
    F_{t+1|t+1} ^TF_{t+1|t+1}  &= \begin{bmatrix} (I - LC)F_{t+1|t}^T  &  L\Gamma_V^T \end{bmatrix} \begin{bmatrix}  F_{t+1|t}(I-LC)^T \\ \Gamma_VL^T \end{bmatrix}, \\
    F_{t+1|t+1} &= \operatorname{qr_r}(F_{t+1|t}(I-LC)^T, \Gamma_VL^T).
\end{align}

\subsection{Final Algorithm}
\begin{algorithm} 
	\begin{algorithmic}[1]
		\caption{Square Root Kalman Filter}\label{alg:mgn}
		\State \textbf{function}  sqkf($\mu_{t|t},F_{t|t},u_t,y_{t+1},A,B,C,\Gamma_W,\Gamma_V)$
		\State 
		\State \quad \# predict step 
		\State \quad $\mu_{t+1|t} = A \mu_{t|t} + B u_t $ \Comment{state prediction}
		\State \quad $F_{t+1|t} = \operatorname{qr_r} ( F_{t|t} A^T , \Gamma_W)$ \Comment{covariance prediction}
		\State 
		\State \quad \# innovation
        \State \quad $z = y_{t+1} - C\mu_{t+1|t}$ \Comment{measurement innovation}
        \State \quad $G = \operatorname{qr_r}(F_{t+1|t} C^T, \Gamma_V)$ \Comment{innovation covariance}
        \State \quad $L = [G^{-1}(G^{-T} C) F_{t+1|t}^TF_{t+1|t}]^T$ \Comment{Kalman gain}
        \State 
        \State \quad \# update step 
        \State \quad $\mu_{t+1|t+1} = \mu_{t+1|t} + Lz $ \Comment{state update}
        \State \quad $F_{t+1|t+1} = \operatorname{qr_r}(F_{t+1|t}(I-LC)^T, \Gamma_VL^T)$ \Comment{covariance update}
        \State 
        \State \quad \textbf{return}($\mu_{t+1|t+1}, F_{t+1|t+1}$)
	\end{algorithmic}
\end{algorithm}

\bibliographystyle{unsrt}  
\bibliography{references}  %%% Remove comment to use the external .bib file (using bibtex).


% George Kour and Raid Saabne.
% \newblock Real-time segmentation of on-line handwritten arabic script.
% \newblock In {\em Frontiers in Handwriting Recognition (ICFHR), 2014 14th
%   International Conference on}, pages 417--422. IEEE, 2014.

% George Kour and Raid Saabne.
% \newblock Fast classification of handwritten on-line arabic characters.
% \newblock In {\em Soft Computing and Pattern Recognition (SoCPaR), 2014 6th
%   International Conference of}, pages 312--318. IEEE, 2014.

% \bibitem{hadash2018estimate}
% Guy Hadash, Einat Kermany, Boaz Carmeli, Ofer Lavi, George Kour, and Alon
% \newblock Estimate and replace: A novel approach to integrating deep neural
%   networks with existing applications.
% \newblock {\em arXiv preprint arXiv:1804.09028}, 2018.

% \end{thebibliography}

\end{document}